\documentclass[11pt,a4paper]{article}

\usepackage[T1]{fontenc}
\usepackage[utf8]{inputenc}
\usepackage{lmodern}

\usepackage[a4paper,margin=1in]{geometry}

\usepackage{amsmath}
\usepackage{amssymb}
\usepackage{amsfonts}
\usepackage{bm}
\usepackage{mathtools}

\usepackage{graphicx}
\usepackage{booktabs}
\usepackage{multirow}
\usepackage{array}
\usepackage{longtable}
\usepackage{tabularx}
\usepackage{caption}
\usepackage{subcaption}
\usepackage{float}
\usepackage{placeins}

\usepackage{siunitx}

\usepackage{xcolor}
\usepackage{hyperref}

\hypersetup{
    colorlinks=true,
    citecolor=blue,
    linkcolor=blue,
    urlcolor=blue
}

\usepackage[numbers,sort&compress]{natbib}

\usepackage{authblk}

\usepackage{titlesec}

\titleformat{\section}
{\Large\bfseries}
{\thesection.}{0.5em}{}

\titleformat{\subsection}
{\large\bfseries}
{\thesubsection.}{0.5em}{}

\usepackage{setspace}
\title{Spectral--Timing Evolution of the Black Hole X-ray Binary LMC X--3: An XPoSat View}

\author[1]{Sayantan Bhattacharya}
\author[1]{Aman Kaushik}
\author[1]{Sudip Bhattacharyya}

\affil[1]{Department of Astronomy and Astrophysics,
Tata Institute of Fundamental Research,
1 Homi Bhabha Road,
Mumbai 400005,
India}

\date{}

\begin{document}

\maketitle

%% Use the \collaboration command to identify collaborations. This command
%% takes an optional argument that is either a number or the word "all"
%% which tells the compiler how many of the authors above the command to
%% show. For example "\collaboration[all]{(DELVE Collaboration)}" wil include
%% all the authors above this command.
%%
%% Mark off the abstract in the ``abstract'' environment. 

\begin{abstract}

LMC X--3 is a persistent black hole high-mass X-ray binary and serves as an important laboratory for studying accretion disk evolution across different spectral states. We present a spectral--timing analysis of \textit{XPoSat}/XSPECT observations of LMC X--3 obtained during 2024 and 2025. Using broadband X-ray light curves, hardness diagnostics, power spectral analysis, and spectral modeling, we investigate the long-term evolution of the source across multiple epochs. LMC X--3 exhibits significant changes in both its spectral and timing properties between the two observing campaigns. Hardness--intensity and hardness--rms diagrams reveal a clear separation between the 2024 and 2025 observations and even within the 2024 observations. Overall, it indicates a transition from a relatively harder and more variable state to a brighter, softer, and less variable state. Power spectral analysis shows enhanced broadband variability during the 2024 observations, while the 2025 observations are characterized by weaker variability. Spectral modeling of the combined XSPECT observations indicates an increase in the thermal disk temperature from 0.8 to 1.17 keV, accompanied by a $\sim$5 times increase in X-ray flux between the two epochs. The observed reduction in fractional rms variability with increasing source brightness and decreasing spectral hardness is consistent with more contribution from a stable thermal accretion disk. Taken together, the spectral and timing diagnostics provide a self-consistent picture of accretion -state evolution in LMC X--3, demonstrating the capability of XSPECT to probe long-term variability and accretion processes in bright black hole X-ray binaries.

\end{abstract}

%% Keywords should appear after the \end{abstract} command. 
%% The AAS Journals now uses Unified Astronomy Thesaurus (UAT) concepts:
%% https://astrothesaurus.org
%% You will be asked to selected these concepts during the submission process
%% but this old "keyword" functionality is maintained in case authors want
%% to include these concepts in their preprints.
%%

\noindent\textbf{Keywords:}
Black hole physics,
X-ray astronomy,
High Energy astrophysics,
High mass x-ray binary stars,
Accretion.
%% From the front matter, we move on to the body of the paper.
%% Sections are demarcated by \section and \subsection, respectively.
%% Observe the use of the LaTeX \label
%% command after the \subsection to give a symbolic KEY to the
%% subsection for cross-referencing in a \ref command.
%% You can use LaTeX's \ref and \label commands to keep track of
%% cross-references to sections, equations, tables, and figures.
%% That way, if you change the order of any elements, LaTeX will
%% automatically renumber them.
\section{Introduction}\label{sec:intro}

Black hole X-ray binaries (BHXBs) provide a unique opportunity to study accretion onto compact objects over a wide range of mass accretion rates and accretion states. These systems exhibit distinct spectral and temporal states that are generally interpreted in terms of changes in the relative contributions of the optically thick accretion disk and the hot comptonizing corona \citep{remillard2006x,dunn2010global}. Transitions between these states are often accompanied by significant changes in X-ray luminosity, spectral hardness, and rapid variability, making simultaneous spectral and timing studies a powerful probe of accretion physics in strong gravitational fields.

The evolution of X-ray variability across accretion states has been extensively studied in Galactic black hole binaries using missions such as \textit{RXTE}, \textit{XMM--Newton}, \textit{NICER}, and \textit{Swift}. Hard states are typically characterized by strong broadband variability and the frequent presence of quasi-periodic oscillations (QPOs), while softer, disk-dominated states exhibit substantially reduced variability amplitudes \citep{done2007modelling}. The connection between spectral properties and rapid X-ray variability provides important constraints on the geometry and evolution of the accretion flow.

LMC X--3 is one of the best-studied persistent black hole X-ray binaries and has long served as a benchmark source for investigations of accretion disk evolution. Located in the Large Magellanic Cloud at a distance of $\sim 50$ kpc, the system consists of a stellar-mass black hole accreting from a B-type companion through Roche-lobe overflow \citep{cowley1984optical,orosz2014mass}. Unlike many transient Galactic BHXBs, LMC X--3 remains persistently active while exhibiting substantial luminosity and spectral variations over timescales ranging from days to years \citep{wilms2001discovery}. Owing to its relatively low absorption column and strong thermal emission, the source has been widely used to investigate accretion disk structure, state transitions, and long-term variability \citep{kubota2010testing, steiner2014low}.

Previous observations have revealed that LMC X--3 spends most of its time in thermally dominated states, although excursions toward harder states have also been reported \citep{nowak2001good,steiner2010constant}. Long-term monitoring campaigns have demonstrated strong correlations between spectral hardness, luminosity, and variability properties, suggesting significant changes in the accretion flow geometry as the mass accretion rate evolves \citep{wilms2001discovery}. Nevertheless, continued monitoring remains valuable for characterizing the source across different epochs and for comparing its spectral and timing behavior using new instrumentation.
\begin{figure*}
\centering
\includegraphics[width=\textwidth]{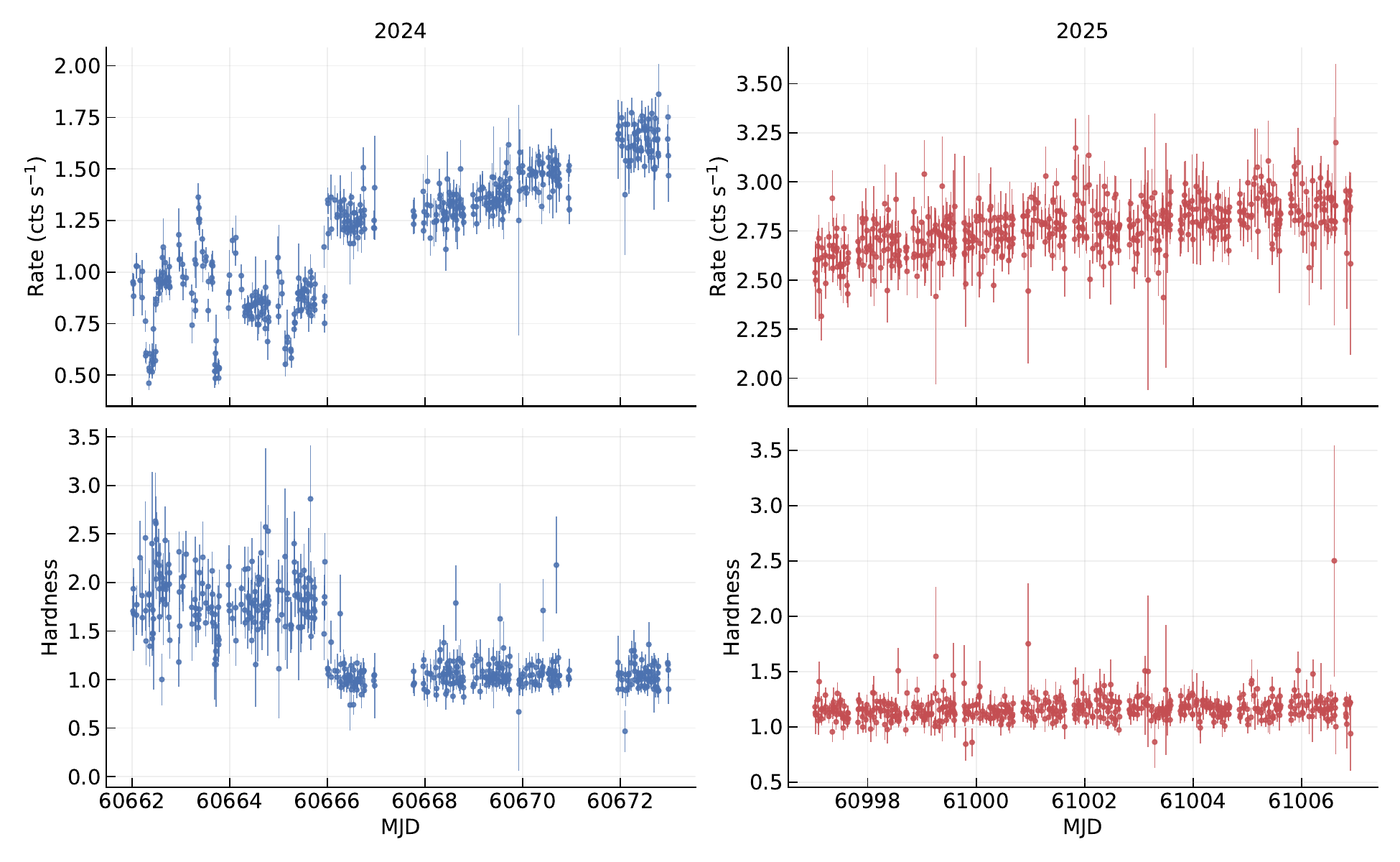}
\caption{Broad-band X-ray light curves (top panels) and hardness ratio evolution (bottom panels) of LMC X--3 observed with \textit{XpoSat/XSPECT} during 2024 (left) and 2025 (right). The source displays substantial spectral evolution between the two epochs, transitioning from a harder and more variable state in 2024 to a softer state in 2025. See Section~\ref{subsec:lc} for details.}
\label{fig:lc_hardness}
\end{figure*}
The \textit{X-ray Polarimeter Satellite} (XPoSat) carries the \textit{X-ray Spectroscopy and Timing} (XSPECT) instrument, which provides simultaneous spectral and timing information in the soft X-ray band. The availability of observations spanning multiple epochs offers an opportunity to investigate the spectral and temporal evolution of bright X-ray binaries such as LMC X--3. In this work, we analyze XSPECT observations obtained during 2024 and 2025 with the goal of characterizing the source evolution through a combination of spectral, timing, and hardness-based diagnostics. By examining the long-term changes in spectral hardness, broadband variability, and X-ray spectral properties, we aim to place the observed behavior within the broader context of accretion state evolution in black hole X-ray binaries. In addition, these observations provide one of the first opportunities to assess the capability of XSPECT for simultaneous spectral and timing studies of bright black hole X-ray binaries over long temporal baselines.

This work presents the first long-term spectral–timing characterization of LMC X-3 with XPoSat/XSPECT, covering two monitoring campaigns separated by approximately one year. This paper is organized as follows. In Section~\ref{sec:intro}, we give an introduction to the topic in general and the motivation for this work. In Section~\ref{sec:obs}, we describe the observations and data reduction procedures. Section~\ref{sec:results} presents the spectral and timing results of LMC X--3. Then the results are discussed in Section~\ref{sec:disc}, and our conclusions are summarized in Section~\ref{sec:conc}.

\section{Observations and Data Analysis}\label{sec:obs}

\subsection{XPoSat/XSPECT}

The \textit{X-ray Polarimeter Satellite} (\textit{XPoSat}) was launched on 2024 January 1 as India's first dedicated X-ray polarimetry mission. In addition to the Polarimeter Instrument in X-rays (POLIX), the mission carries the \textit{X-ray SPECtroscopy and Timing} (XSPECT) payload, which provides simultaneous spectroscopic and timing observations in the 0.8--15 keV energy range \citep{vatedka2025x}. XSPECT consists of an array of Swept Charge Devices (SCDs) arranged in four detector quadrants, providing fields of view of $2^\circ\times2^\circ$ and $3^\circ\times3^\circ$. The instrument offers an energy resolution of approximately 180 eV at 6 keV and a fast readout time of $\sim1$ ms, enabling timing studies of bright X-ray sources while minimizing pile-up effects \citep{chatterjee2025xspect,prakash2025spectral,vatedka2025x}.

\begin{table}
\centering
\caption{{\it XPoSat}/XSPECT observations of LMC~X--3 used in this paper.}
\label{tab:obslog}
\begin{tabular}{lccr}
\hline
ObsID & Observation Date & Exposure (ks) \\
\hline

\multicolumn{3}{c}{\textbf{2024}}\\
\hline
T24\_0013 & 2024-12-18 & 18.94 \\
-- & 2024-12-19 & 19.31 \\
-- & 2024-12-20 & 18.50 \\
-- & 2024-12-21 & 17.66 \\
T24\_0012 & 2024-12-22 & 19.78 \\
-- & 2024-12-23$^{\dagger}$ & 2.95 \\
-- & 2024-12-24 & 20.88 \\
-- & 2024-12-25 & 20.29 \\
-- & 2024-12-26 & 18.55 \\
-- & 2024-12-27$^{\dagger}$ & 1.44 \\
-- & 2024-12-28 & 22.09 \\

\hline
\multicolumn{3}{c}{\textbf{2025}}\\
\hline
T25\_0021 & 2025-11-18 & 20.67 \\
-- & 2025-11-19 & 20.16 \\
-- & 2025-11-20 & 18.93 \\
-- & 2025-11-21 & 19.38 \\
-- & 2025-11-22 & 19.59 \\
-- & 2025-11-23 & 18.91 \\
-- & 2025-11-24 & 18.68 \\
-- & 2025-11-25 & 18.02 \\
-- & 2025-11-26 & 18.05 \\
-- & 2025-11-27 & 17.02 \\

\hline
\end{tabular}

\vspace{2mm}
\begin{minipage}{0.95\linewidth}
\footnotesize
$^{\dagger}$ These observations have significantly shorter effective exposures than the remaining observations and were excluded from the merged spectral analysis due to insufficient photon statistics.
\end{minipage}

\end{table}

\subsection{Observations}

LMC X--3 was observed by XSPECT during two monitoring campaigns in 2024 and 2025. The observations span multiple epochs and sample distinct spectral states of the source. All publicly available Level-2 XSPECT products corresponding to these observations were downloaded from the Indian Space Science Data Centre (ISSDC) archive.

For the present study, we restricted our analysis to the $2^\circ\times2^\circ$ field-of-view detectors in order to maintain a homogeneous data set across all observations. The Level-2 event files, light curves, spectra, response matrices, and ancillary response files generated by the XSPECT standard processing pipeline were used throughout this work.

\subsection{Data Reduction}

The XSPECT data were processed using the standard mission pipeline products available through ISSDC. The archival Level-2 event files, which already incorporate the recommended screening criteria and Good Time Intervals (GTIs), were adopted for all subsequent analysis. Additionally the event files were filtered from flares using \texttt{xspfilterflares} command.

The observations consisted of multiple individual exposures in each year, the Level-2 event files were merged using the \texttt{xspevtmerge} task to create epoch-averaged event files for spectral analysis. Source spectra were generated using the standard XSPECT spectral extraction command (\texttt{xspspecgen}) and the corresponding response matrix files (RMFs) (provided from the XSPECT archive) and ancillary response files (ARFs) generated by the pipeline. For the timing analysis, light curves were extracted directly from the Level-2 event files in multiple energy bands (\texttt{xsplcgen} command).

To investigate spectral evolution, broad (0.8--10.0 keV), soft (0.8--2.0 keV), and hard (2.0--10.0 keV) light curves (500s time bin) were produced using the XSPECT pipeline. Hardness ratios were defined as the ratio of the hard-band to soft-band count rates. Hardness--intensity and hardness--rms diagrams were constructed using these light curves after removing statistically significant outliers through sigma-clipping.

For spectral analysis, all available observations from a given epoch were combined to improve photon statistics. Spectral fitting was performed in XSPEC version 12.14 \citep{arnaud1999xspec}. The spectra were grouped to ensure sufficient counts per bin for $\chi^2$ minimization. Unless otherwise noted, all uncertainties quoted in this work correspond to the 68 per cent confidence interval for a single interesting parameter.

For timing studies, power density spectra were computed from the broadband light curves and normalized following the fractional rms prescription of \citet{belloni2002unified}. Hardness-dependent variability was further quantified using hardness--rms diagrams. We additionally searched for inter-band delays between the soft and hard energy bands using cross-correlation of light curves binned at 10 milliseconds.

\section{Results}\label{sec:results}

\subsection{Long-Term X-ray Variability}\label{subsec:lc}

Figure~\ref{fig:lc_hardness} shows the broad-band X-ray light curves and hardness evolution of LMC X--3 during the 2024 and 2025 observing campaigns. The source exhibits substantial variability on timescales of days in both epochs. The 2024 observations show a broader range of count rates and hardness values, whereas the 2025 observations remain confined to a relatively narrow region of parameter space.

The hardness evolution reveals a systematic difference between the two epochs. During 2024, LMC X--3 occupied a harder state characterized by large excursions in hardness ratio. In contrast, the 2025 observations are clustered around lower hardness values, indicating a softer spectral state. The reduced scatter in hardness during 2025 suggests a more stable accretion configuration compared to 2024.

\subsection{Hardness--Intensity Evolution}

\begin{figure*}
\centering
\includegraphics[width=\textwidth]{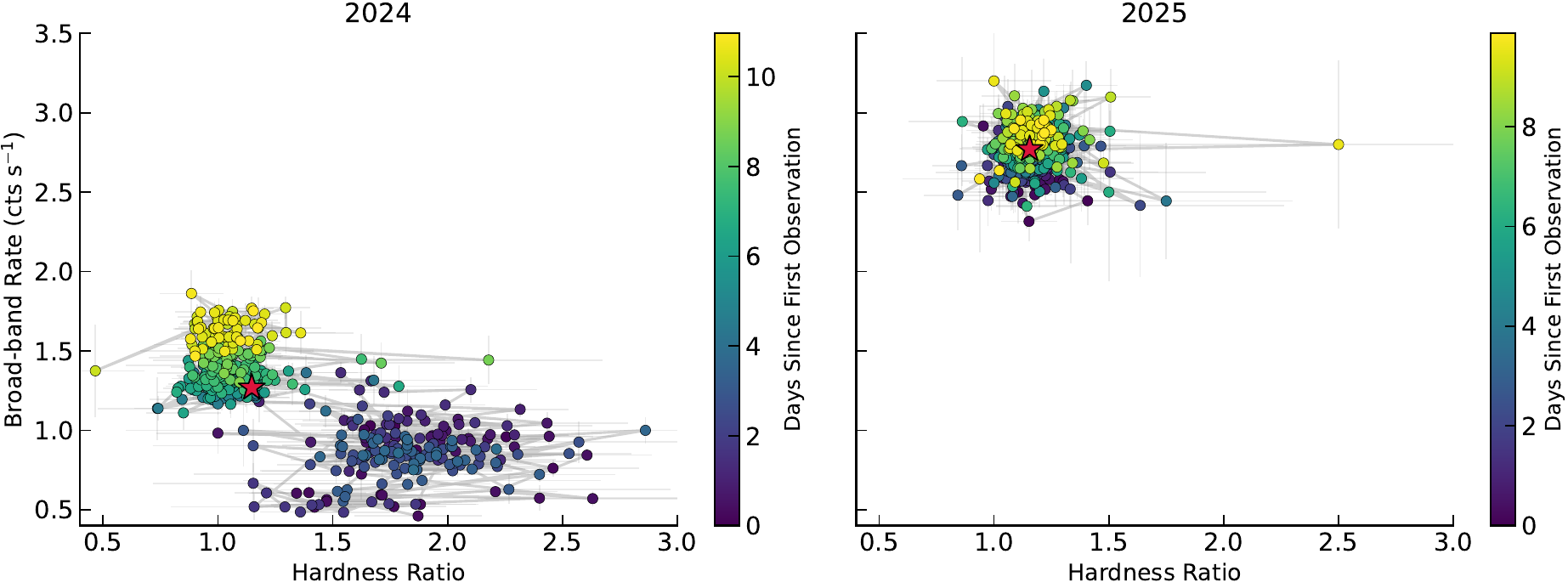}
\caption{Hardness--intensity diagrams of LMC X--3 for the 2024 and 2025 XSPECT/XpoSat observing campaigns. Data points are colour-coded by observation time. The source occupies a significantly harder and more extended region of the HID during 2024, while the 2025 observations cluster in a softer and brighter state.}
\label{fig:hid}
\end{figure*}

The hardness--intensity diagrams (HIDs) for the two epochs are shown in Figure~\ref{fig:hid}. Each point represents an individual time bin and is colour-coded by observation time. The HID reveals a clear separation between the source populations sampled in 2024 and 2025.

The 2024 observations span a wide range of hardness values and intensities, indicating substantial spectral evolution during the observing campaign. The source occupies both relatively hard and intermediate spectral states. In contrast, the 2025 observations are concentrated at lower hardness ratios and higher count rates, consistent with a thermally dominated soft state.

The time evolution encoded in the color scale demonstrates a gradual migration of the source through the HID during the 2024 observations, while the 2025 observations remain comparatively stable.

\subsection{Hardness--RMS Relation}

\begin{figure}
\centering
\includegraphics[width=0.5\columnwidth]{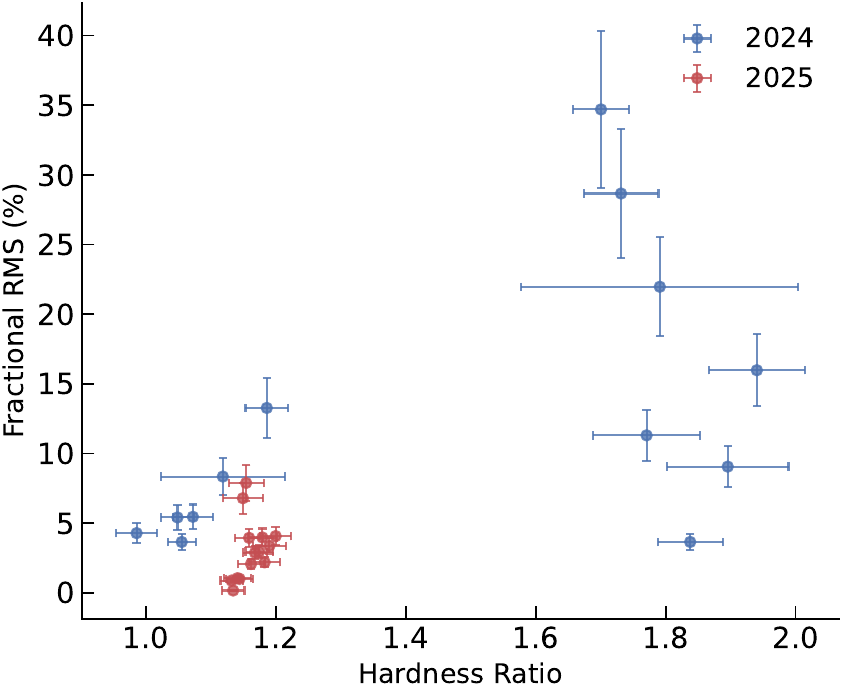}
\caption{Hardness--rms diagram of LMC X--3. The fractional rms variability decreases systematically toward softer spectral states observed by XSPECT/XpoSat. The 2025 observations occupy a low-rms region compared to the harder 2024 observations, which span a large range of rms values.}
\label{fig:hrd}
\end{figure}

To investigate the connection between spectral state and rapid variability, we constructed hardness--rms diagrams (HRDs) using the segmented light curves. The resulting HRD is presented in Figure~\ref{fig:hrd}.

The fractional rms variability is strongly dependent on spectral hardness. The hardest observations display the largest variability amplitudes, whereas softer observations exhibit substantially reduced rms values. The 2024 data populate the high-hardness and high-variability region of the diagram, while the 2025 observations are confined to lower hardness ratios and lower rms amplitudes.

This behaviour is consistent with the well-established spectral--timing evolution observed in black hole X-ray binaries, where increasing thermal disk dominance is accompanied by suppression of rapid aperiodic variability. The positive correlation between spectral hardness and fractional rms variability indicates that the harder observations are associated with stronger aperiodic variability, while the softer observations exhibit suppressed broadband variability.

\subsection{Broad-band Timing Properties}

\begin{figure}
\centering
\includegraphics[width=0.5\columnwidth]{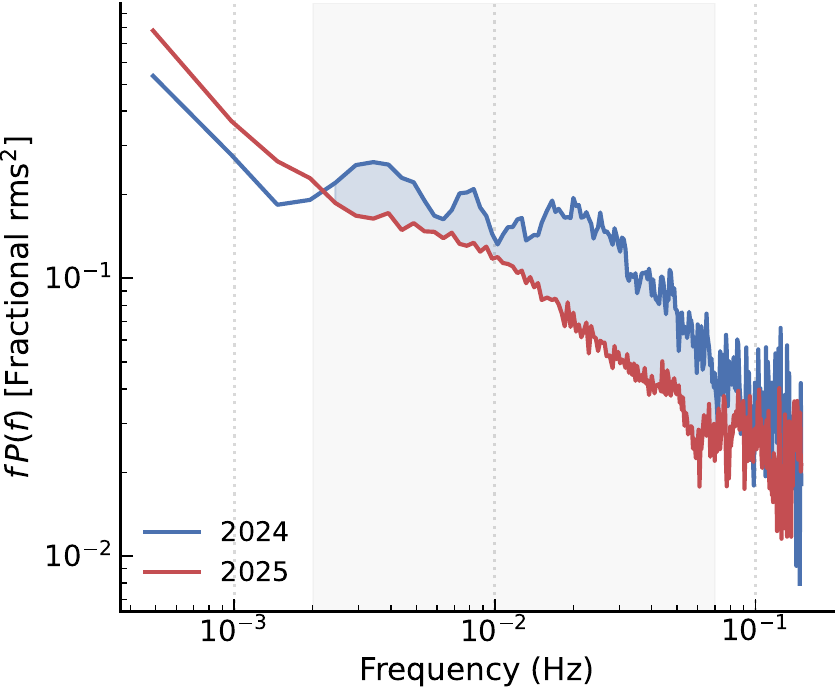}
\caption{Broad-band power density spectra of LMC X--3 during 2024 and 2025 XSPECT/XpoSat observations. The harder 2024 state exhibits enhanced low-frequency variability relative to the softer 2025 observations. No significant quasi-periodic oscillations are detected.}
\label{fig:pds}
\end{figure}

The broad-band power density spectra (PDS) derived from the 2024 and 2025 observations are shown in Figure~\ref{fig:pds}. Both epochs are dominated by red-noise variability with power increasing toward lower frequencies.

The 2024 observations exhibit significantly enhanced variability power over a broad frequency range compared to 2025. Excess low-frequency variability is particularly evident below $\sim 0.05$ Hz. No statistically significant quasi-periodic oscillations are detected in either epoch.

The suppression of broad-band variability during 2025 is consistent with the softer spectral state inferred from the HID and HRD analyses. Together, these diagnostics suggest an increasing contribution from a stable thermal accretion disk.

\subsection{Spectral Analysis}

The combined XSPECT spectra obtained during the 2024 and 2025 campaigns were fitted using an absorbed multicolour disk blackbody model. For 2024, only the soft duration is fitted as the harder segment was fainter and for less duration, hence lacking statistics. The best-fitting spectral parameters are summarized in Table~\ref{tab:specfit}.

The spectra are adequately described by the model
[
$\mathrm{tbabs \times diskbb}$
]
with acceptable fit statistics in both epochs. We also tested tbabs$\times$powerlaw, tbabs$\times$(diskbb+powerlaw), but the inclusion of the powerlaw component gave a poorer fit. The spectral fits indicate a significant increase in the inner disk temperature between 2024 and 2025, accompanied by an increase in source flux. The increase in disk temperature together with the reduction in hardness and variability indicates that the source evolved toward a more thermally dominated accretion state during 2025.

\begin{figure}
\centering
\includegraphics[width=0.5\columnwidth]{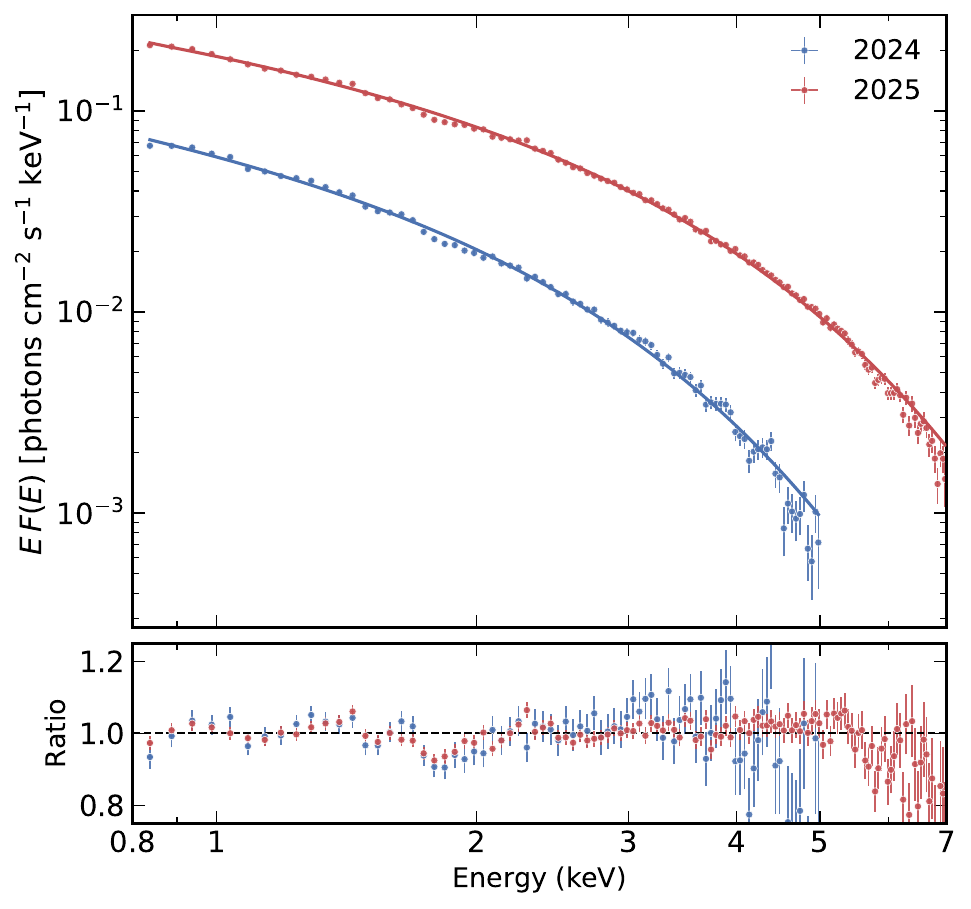}
\caption{XSPECT/XpoSat spectra of LMC X--3 obtained during the 2024 (soft only) and 2025 observing campaigns together with the best-fitting absorbed multicolour disk blackbody model (tbabs $\times$ diskbb). Ratios are shown in the lower panel.}
\label{fig:spectra}
\end{figure}

\begin{table*}
\centering
\caption{Best-fitting spectral parameters of LMC~X--3 obtained from the XPoSat/XSPECT spectra (Section~3.5).}
\label{tab:specfit}
\begin{tabular}{lcc}
\hline
Parameter & 2024 & 2025 \\
\hline
$N_{\rm H}$ ($10^{22}$ cm$^{-2}$) & 0.01$^{\dagger}$ & 0.011$\pm$0.004 \\
$kT_{\rm in}$ (keV) & 0.837$\pm$0.005 & 1.167$\pm$0.004 \\
Norm$_{\rm disk}^{\S}$ & 24.8$\pm$0.7 & 29.1$\pm$0.5 \\
Unabsorbed Flux ($10^{-10}$ erg cm$^{-2}$ s$^{-1}$) & 1.964$\pm$0.002 & 9.585$\pm$0.004 \\
$\chi^{2}$ (dof) & 381.13 (331) & 459.38 (494) \\
\hline
\end{tabular}

\vspace{2mm}
\footnotesize
$^{\dagger}$ Hydrogen column density was frozen during the fit because it could not be constrained.\\
$^{\ddagger}$ The 2024 spectrum corresponds to the combined \emph{soft}-state observations only.\\
$^{\S}$ Norm$_{\rm disk}=\left(\frac{R_{\rm in}/\mathrm{km}}{D/10~\mathrm{kpc}}\right)^2\cos i$, where $R_{\rm in}$ is the apparent inner disk radius, $D$ is the source distance, and $i$ is the disk inclination.\\
$^{\P}$ Uncertainties are quoted at the 68\% confidence level for one interesting parameter.

\end{table*}

\section{Discussion}\label{sec:disc}

The XPoSat/XSPECT monitoring presented here provides a consistent picture of the spectral and timing evolution of LMC~X--3 across two observing campaigns separated by approximately one year. The source evolves from a relatively harder, more variable, and fainter state during the 2024 observations to a brighter, softer, and more thermally dominated state during the 2025 campaign. In addition to the clear separation between the two epochs, the hardness--intensity diagram demonstrates a gradual evolution within the 2024 observations themselves, indicating that the source continuously migrated toward softer and brighter emission during the course of the monitoring. Such behaviour is characteristic of changes in the accretion flow associated with increasing mass accretion rate in black hole X-ray binaries \citep{remillard2006x,done2007modelling}.

The observed timing behaviour is fully consistent with this spectral evolution. The hardness--rms diagram reveals that observations with larger hardness ratios exhibit systematically stronger fractional rms variability, whereas softer observations are characterized by substantially weaker variability. This positive correlation between spectral hardness and rapid X-ray variability is a well-established property of black hole X-ray binaries and is generally interpreted as a consequence of the changing relative contributions of the optically thick accretion disk and the variable hard X-ray emitting region \citep{belloni2002unified}. As the thermal disk contributes an increasing fraction of the observed X-ray emission, the intrinsically stronger variability associated with the hot inner flow becomes progressively diluted, leading to lower fractional rms amplitudes\citep{done2007modelling}. The power density spectra support this interpretation by showing enhanced low-frequency broadband variability during the harder 2024 observations and a significant suppression of variability during the softer 2025 epoch. No statistically significant quasi-periodic oscillations or inter-band time lags were detected in either campaign, consistent with the absence of a dominant rapidly varying hard spectral component within the XSPECT energy band.

The spectral evolution independently supports the same physical picture. Both epochs are adequately described by the absorbed multicolour disk blackbody model \texttt{tbabs$\times$diskbb}, while the addition of an extra Comptonization component does not produce a statistically significant improvement to the spectral fits. Such behaviour is consistent with previous studies showing that LMC~X--3 spends the majority of its lifetime in thermally dominated states, with the X-ray spectrum largely governed by emission from an optically thick, geometrically thin accretion disk \citep{wilms2001discovery,nowak2001good,kubota2010testing,steiner2010constant}. The increase in the inner disk temperature from $kT_{\rm in}\approx0.84$~keV in 2024 to $1.17$~keV in 2025, accompanied by an approximately fivefold increase in X-ray flux, indicates that the accretion disk became both hotter and more luminous during the later observations. These values are consistent with the range of disk temperatures previously reported for LMC~X--3 during its thermally dominated state \citep{kubota2010testing,steiner2014low}, suggesting that both observing campaigns sampled disk-dominated accretion, with the 2025 observations representing a hotter and brighter thermal state.

The fitted \texttt{diskbb} normalization provides a useful consistency check on the inferred accretion geometry. Assuming a source distance of $\sim50$~kpc and a binary inclination of $\sim67^\circ$, the best-fitting normalization for the 2025 spectrum corresponds to an apparent inner disk radius of approximately 43~km. Applying the standard corrections for spectral hardening and the stress-free inner boundary condition \citep{kubota1998} yields a physical radius of approximately 50~km. For a black hole mass of $\sim7\,M_\odot$ and a dimensionless spin parameter of $a_\ast\approx0.25$, the expected radius of the innermost stable circular orbit (ISCO) is approximately 53~km\citep{orosz2014mass,steiner2014low}. Although the \texttt{diskbb} normalization alone cannot provide a precise measurement of the inner disk radius because of uncertainties in the colour-correction factor, inclination, and relativistic effects, the inferred radius is nevertheless consistent with previous continuum-fitting studies indicating that the accretion disk in LMC~X--3 remains close to the ISCO during thermally dominated states \citep{steiner2010constant,steiner2014low}. The present analysis therefore does not provide an independent measurement of the ISCO location, but rather demonstrates that the XSPECT observations are fully consistent with the established accretion geometry of the source.

Our results are broadly consistent with more than two decades of observations of LMC~X--3 obtained with missions such as \textit{RXTE}, \textit{XMM--Newton}, and \textit{NICER}, which have established the source as a prototype of a persistently accreting, disk-dominated black hole X-ray binary exhibiting long-term spectral and luminosity variations \citep{wilms2001discovery,kubota2010testing}. While the present observations do not reveal previously unknown accretion states, they provide the first long-term spectral--timing characterization of LMC~X--3 with XPoSat/XSPECT and demonstrate that the instrument successfully reproduces the canonical relationships between spectral hardness, rapid X-ray variability, and thermal disk evolution. These results establish XSPECT as a capable facility for future long-term monitoring of bright Galactic X-ray binaries and for investigating accretion-state evolution using simultaneous spectral and timing diagnostics.

\section{Conclusions}\label{sec:conc}

We have presented a comprehensive spectral--timing analysis of LMC~X--3 using {\it XPoSat}/XSPECT observations obtained during the 2024 and 2025 monitoring campaigns. By combining hardness diagnostics, timing analysis, and X-ray spectral modeling, we characterize the long-term evolution of the source and demonstrate the capability of XSPECT to investigate accretion-state evolution in bright black hole X-ray binaries.

Our principal results are summarized as follows:

\begin{itemize}

\item The broad-band light curves and hardness evolution reveal significant long-term spectral evolution, with the 2024 observations sampling a harder and more variable state, while the 2025 observations correspond to a brighter, softer, and more stable accretion state.

\item The hardness--intensity diagram shows a clear separation between the two observing epochs, while the 2024 observations themselves trace a gradual evolution from harder to softer states over the course of the monitoring campaign.

\item The hardness--rms diagram demonstrates that harder observations exhibit larger fractional rms variability systematically, whereas the softer observations show substantially reduced variability, consistent with an increasing contribution from a stable thermal accretion disk.

\item Broad-band power density spectra reveal enhanced low-frequency variability during the 2024 observations, while the variability is significantly suppressed in 2025. No statistically significant quasi-periodic oscillations or inter-band time lags were detected in either epoch.

\item Spectral fitting with an absorbed multicolor disk blackbody model indicates an increase in the inner disk temperature from $kT_{\rm in}\approx0.84$~keV in 2024 to $\approx1.17$~keV in 2025, accompanied by an increase in the X-ray flux by nearly a factor of five. The inferred inner disk radius is consistent with the accretion disk extending close to the innermost stable circular orbit during the thermally dominated state.

\end{itemize}

Taken together, the spectral and timing diagnostics present a consistent picture of LMC~X--3 evolving toward a disk-dominated accretion state characterized by increasing thermal emission and suppressed rapid variability. These results are in good agreement with the established phenomenology of black hole X-ray binaries and demonstrate that XSPECT is capable of simultaneously probing the spectral and timing evolution of bright accreting compact objects over long-term monitoring campaigns.

%%%%%%%%%%%%%%%%%%%%%%%%%%%%%%%%%%%%%%%%%%%%%%%%%%
\section*{Acknowledgements}

This research has made use of XPoSat data obtained through the Indian Space Science Data Centre (ISSDC), ISRO. We acknowledge the support received from TIFR Mumbai to carry out this research. The usage of AI has been limited to editing and rephrasing the document; AI has not been used for any creative purposes. 

%%%%%%%%%%%%%%%%%%%%%%%%%%%%%%%%%%%%%%%%%%%%%%%%%%%%%%%%%%%%%%%%%%%%%%%%%%
\section*{Data Availability}

The XPoSat/XSPECT data analysed in this work are publicly available through the Indian Space Science Data Centre (ISSDC). The processed data products and analysis scripts used in this work are available from the corresponding author upon reasonable request.

%%%%%%%%%%%%%%%%%%%%%%%%%%%%%%%%%%%%%%%%%%%%%%%%%%%%%%%%%%%%%%%%%%%%%%%%%%

\bibliographystyle{aasjournal}
\bibliography{references}

\end{document}